\documentclass[pra,10pt,notitlepage,showkeys]{revtex4-1}

\usepackage{amsmath}

\DeclareMathOperator{\Tr}{Tr}
\DeclareMathOperator{\tr}{tr}

\begin{document}

\title{Derivation of the effective action of a dilute Fermi gas in the
  unitary limit of the BCS-BEC crossover}

\author{Adriaan M. J.~Schakel}

\affiliation{Institut f\"ur Theoretische Physik, Universit\"at Leipzig,
  Postfach 100 920, D-04009 Leipzig, Germany}
\affiliation{
Laborat\'orio de F\'isica Te\'orica e Computacional,
Departamento de F\'{i}sica,
Universidade Federal de Pernambuco, 
50670-901, Recife-PE, Brazil\footnote{Present address}}

\begin{abstract}
  The effective action describing the gapless Nambu-Goldstone, or
  Anderson-Bogoliubov, mode of a zero-temperature dilute Fermi gas at
  unitarity is derived up to next-to-leading order in derivatives from
  the microscopic theory.  Apart from a next-to-leading order term that
  is suppressed in the BCS limit, the effective action obtained in the
  strong-coupling unitary limit is proportional to that obtained in the
  weak-coupling BCS limit.  
\end{abstract}

\maketitle

\section{Introduction}
The ability to control the interaction between atoms is unique to
experiments on optically trapped ultracold atomic gases, unmatched by
other condensed matter experiments and unthought of even a few years
ago.  After successfully creating a degenerate Fermi sea in a trapped
Fermi gas \cite{DeMarco:1999}, Jin and collaborators were able to tune a
dilute ultracold Fermi gas to be near a magnetic Feshbach resonance
where small changes in magnetic field strength have pronounced effects
on the two-particle scattering length characterizing the interatomic
pair potential in vacuum \cite{Regal:2003}.  Using this handle, the
group produced in 2003 for the first time a condensate of pairs of
fermionic atoms outside the weak-coupling Bardeen-Cooper-Schrieffer
(BCS) limit of loosely bound Cooper pairs \cite{Greiner:2003}, and
subsequently managed to realize the intriguing crossover from the BCS
limit to the limit of tightly bound pairs that form a Bose-Einstein
condensate\cite{Regal:2004}.  This crossover, which was studied earlier
by a number of
theorists~\cite{Eagles:1969,Leggett:1980,Nozieres:1985,Randeria:1990,Drechsler:1992,Engelbrecht:1997,Marini:1998},
is accompanied by a smooth change in chemical potential from positive in
the BCS limit to negative in the Bose-Einstein condensate (BEC) limit.

The unitary, or strong-coupling, limit of the BCS-BEC crossover, which
is in the region with positive chemical potential, is of particular
interest \cite{Bertsch:1998}.  This limit marks the threshold of a bound
state in vacuum, where the scattering length diverges and changes from
negative on the BCS side to positive on the BEC side.  Because the
scattering length diverges in this limit, the dilute system has no
obvious scale parameter, other than the particle number density $n$.
This implies, for example, the remarkable result that the ground-state
energy of this strongly interacting system is proportional to that of a
free Fermi gas, or equivalently, a neutral BCS superconductor at the
same density.  The absence of a coupling constant, on the one hand,
precludes a standard perturbative approach, but on the other makes
feasible an \textit{effective field theory approach}.  This is because,
as for critical phenomena characterized by a diverging length scale, the
absence of an intrinsic scale gives rise to universal behavior.  The
effective field theory program differs from perturbation theory in that
it is carried out not by expanding in a small interaction-related
parameter, but by expanding in powers of energy and momentum instead.
Using general coordinate and conformal invariance as guiding principle,
Son and Wingate~\cite{Son:2006}, in an original paper, proposed the most
general effective Lagrangian to leading and next-to-leading orders in a
gradient, or momentum, expansion.  The conformal transformations involve
arbitrary reparametrizations of time $t \to t'(t)$, while the general
coordinate transformations are restricted to curved space (as opposed to
spacetime).  By keeping the metric $g_{ij}$ ($i,j=1,2,3$) arbitrary in
intermediate steps, they were able to derive nontrivial results that
survive the limit of flat space.  At leading order, these new symmetry
arguments are not more powerful than those based on just Galilei
invariance, but they are claimed to be more powerful at the
next-to-leading order \cite{Son:2006}.

The importance of Galilei invariance in describing (clean) BCS
superconductors at the absolute zero of temperature was already stressed
in the 1960ies by Kemoklidze and Pitaevskii \cite{Pitaevskii:1966},
following a suggestion by Nozi\`eres.  It has been used as a guiding
principle for obtaining effective theories of zero-temperature neutral
superconductors as well as superfluids by others since
\cite{Popov:1972,Popov:1976,Takahashi:1988,Greiter:1989}.  Such
effective theories are expressed in terms of the Nambu-Goldstone mode
emerging from the spontaneous breakdown of the global U(1) phase
symmetry in such systems.  Since this mode is gapless, it constitutes
the most important degree of freedom at low frequencies and long wave
lengths.  Its existence in neutral BCS superconductors was first pointed
out by Anderson~\cite{Anderson:1958} and
Bogoliubov~\cite{Bogoliubov:1958a}.  Invariance under global phase
transformations implies that a Nambu-Goldstone field is invariably
accompanied by at least one derivative.  Galilei invariance then
restricts how gradient terms can appear in combination with terms
involving time derivatives.

In this paper, the effective field theory (EFT) describing a dilute
Fermi gas at unitarity and at the absolute zero of temperature up to
next-to-leading order is derived from the microscopic theory using a
derivative expansion method.  This EFT program was first carried out in
Ref.~\cite{effBCS} in the weak-coupling BCS limit and in
Ref.~\cite{effbos} for a weakly interacting Bose gas, and is extended
here to the strong-coupling unitary limit, for details see the
textbook~\cite{BBS}.  Due to a vanishing vertex and conspiring
contributions, coefficients in the EFT at unitarity can, against common
beliefs, be computed analytically beyond the Gaussian
approximation---albeit approximately.  By construction, the approach by
Son and Wingate \cite{Son:2006} is limited to an expansion in momentum
(or inverse wave length).  Since the derivative expansion method
\cite{Fraser:1984zb,Aitchison:1985pp} we adopt treats time and spatial
derivatives on equal footing, it does not face this limitation and
yields an expansion in both momentum and energy (or frequency).  This
leads to two additional terms in the EFT at next-to-leading order
omitted in Ref.~\cite{Son:2006}.  As far as static response functions
are concerned, these additional terms are immaterial.  But they are
relevant for the spectrum of the Nambu-Goldstone mode, or phonons, at
next-to-leading order and modify the result obtained in
Ref.~\cite{Son:2006}.

The paper is organized as follows.  The next section sets the stage and
briefly introduces the derivative expansion method we adopt.
Sections~\ref{sec:LO} and \ref{sec:NLO} treat the effective theory at
leading and next-to-leading order, respectively.  Section~\ref{sec:1D}
discusses the one-dimensional case for which exact results are known.
Finally, integrals and vertices needed in this study are collected in
the Appendix.

For notational convenience, we adopt a relativistic notation.  A
spacetime point will be indicated by the four-vector $x = x^{\mu} = (t,
{\bf x}), \; \mu = 0,1,2,3$, while the frequency $\omega$ and momentum
${\bf k}$ of a particle will be denoted by $k^\mu = (\omega,{\bf k})$.
The time derivative $\partial_t = \partial/\partial t$ and the gradient
$\nabla$ are combined in a single vector $\partial_\mu = (\partial_t,
\nabla)$.  Indices are raised and lowered with the help of the diagonal
metric $\eta_{\mu \nu} = \mathrm{diag}(1,-1,-1,-1)$, so that, for example,
$\partial^\mu = (\partial_t, -\nabla)$.  We also write $k \cdot x =
k_\mu x^\mu = k^\mu x_\mu$ for $\omega t - {\bf k} \cdot {\bf x}$, and
use Einstein's summation convention.  Natural units $\hbar = c = 1$ are
adopted throughout.

\section{Derivative expansion}
A dilute Fermi gas at the absolute zero of temperature can be modeled by
the Lagrangian density
\begin{equation}
\label{31}
     {\cal L} = \sum_{\sigma=\uparrow, \downarrow} \psi^\ast_\sigma
\left(\mathrm{i} \frac{\partial}{\partial t} + \frac{\nabla^2}{2 m}
 + \mu\right) \psi_\sigma - \lambda \psi_{\uparrow}^{\ast}
\psi_{\downarrow}^{\ast} \psi_{\downarrow} \psi_{\uparrow} ,
\end{equation}
where $\psi_{\uparrow (\downarrow )}$ is an anti-commuting field that
describes the fermionic atoms of mass $m$ and spin up (down), and $\mu$
is the chemical potential.  The true interatomic pair potential, which
has typically a repulsive hard core of radius less than one nanometer
and a weak long-range attractive tail, is approximated by a local, i.e.,
delta-function potential, characterized by a single interaction
parameter $\lambda$.  This parameter is related to the (s-wave)
scattering length $a$ of the true potential, parameterizing 
two-particle scattering at low energy in vacuum, through
\begin{equation}
\label{lar}
\frac{1}{\lambda} = \frac{\Gamma({D/2-1})}{4 \pi^{D/2}}
  \frac{m}{a^{D-2}}
\end{equation}
in dimensional regularization.  For later convenience, we have recorded
the result for arbitrary number of space dimensions $D$, and $\Gamma$
denotes the gamma function.  For $D=3$, Eq.~(\ref{lar}) reduces to
$1/\lambda = m/4 \pi a$.  The Lagrangian (\ref{31}) is invariant under
Galilei transformations, where it is recalled that under a Galilei boost
with a constant velocity ${\bf u}$, the coordinates transform as
\begin{equation}
  \label{bs:boost}
t \rightarrow t' = t, \quad {\bf x} \rightarrow {\bf x}' = {\bf x} -
{\bf u} t
\end{equation}
so that
\begin{equation}
\label{bs:boostp}
\frac{\partial}{\partial t} \rightarrow \frac{\partial}{\partial t'} =
\frac{\partial t}{\partial t'} \frac{\partial }{\partial t} +
\frac{\partial \mathbf{x}}{\partial t'} \cdot \nabla = \partial_t +
{\bf u} \cdot \nabla, \quad \nabla \rightarrow \nabla' = \nabla,
\end{equation}
and $\psi(x)$ picks up an extra phase factor
\begin{equation}
\label{bs:phiboost}
\psi(x) \rightarrow \psi'(x') = {\rm e}^{\mathrm{i}m (- {\bf u}
    \cdot {\bf x} + \frac{1}{2}  {\bf u}^2 t) } \, \psi (x) .
\end{equation}

After a Hubbard-Stratonovich transformation, the zero-temperature
partition function,
\begin{equation}     
\label{bcs:34}
Z = \int \mathrm{D} \Psi^{\dagger} \mathrm{D} \Psi \exp \left(
\mathrm{i} \int \mathrm{d}^4 x \, {\cal L} \right),
\end{equation}
can be written in the standard form quadratic in the fermion fields at
the expense of additional integrals over auxiliary fields $\Delta$ and
$\Delta^\dagger$:
\begin{eqnarray} 
\label{bcs:36}
\lefteqn{Z = \int \mathrm{D} \Psi^{\dagger} \mathrm{D} \Psi \int
  \mathrm{D} \Delta^* \mathrm{D} \Delta \; \exp\left(\mathrm{i}
  \frac{1}{\lambda} \int \mathrm{d}^4 x \, |\Delta(x)|^2 \right)}
\nonumber \\ & & \times \exp \left[\mathrm{i} \int \mathrm{d}^4 x \,
  \Psi^{\dagger} \left(
\begin{array}{cc} \mathrm{i}  \partial_t +  \nabla^2/2m + \mu
 & -\Delta(x) \\ -\Delta^*(x) & \mathrm{i}  \partial_t - 
\nabla^2/2m - \mu
\end{array} \right) \Psi \right] , \nonumber  \\ &&
\end{eqnarray} 
where $\Psi$ stands for the two-component Nambu spinor
\begin{equation} \label{32}
\Psi \equiv \left( \begin{array}{c} \psi_{\uparrow} \\
           \psi_{\downarrow}^{\ast}  \end{array} \right),  \quad
    \Psi^{\dagger} = (\psi_{\uparrow}^{\ast},\psi_{\downarrow}) .
\end{equation}
The fermionic degrees of freedom can now be integrated out exactly with
the result
\begin{equation} 
\label{37} 
Z = \int \mathrm{D} \Delta^* \mathrm{D}
\Delta \, \mathrm{e}^{\mathrm{i} S[ \Delta^*, \Delta]},
\end{equation}
where $S[ \Delta^*, \Delta]$ denotes the one-fermion-loop effective
action
\begin{equation}  
\label{312}
S[\Delta^*, \Delta] \equiv \frac{1}{\lambda} \int \mathrm{d}^4 x
 \, |\Delta(x)|^2 - \mathrm{i}  \Tr \ln \left(
\begin{array}{cc} p_0 - \xi ({\bf p}) & -\Delta(x) \\
-\Delta^*(x) & p_{0} + \xi ({\bf p}) \end{array}\right) ,
\end{equation}
with $\xi(\mathbf{k}) = \mathbf{k}^2/2m - \mu$ the kinetic energy of
noninteracting fermions measured relative to the chemical potential.
The trace appearing here is evaluated by using plane waves as a basis
\begin{equation} 
\label{Sgeneff}  
\Tr \ln \left\{ K[p,\Delta(x)] \right\} = \tr \int \mathrm{d}^4 x \int
\frac{\mathrm{d}^4 k}{(2 \pi)^4} \, \mathrm{e}^{\mathrm{i} k \cdot x} \,
\ln \left\{K[p,\Delta(x] \right\} \mbox{e}^{-\mathrm{i} k \cdot x} ,
\end{equation}
with ``tr'' denoting the trace over the discrete indices.  We adopt
the convention that the momentum operator $p^\mu = (\mathrm{i}
\partial_t, -\mathrm{i} \nabla)$ acts on all the fields to its right,
whereas the ordinary derivative $\partial^{\mu}=(\partial_t,-\nabla)$ acts
only on the next field to its right.  The integral $\int \mathrm{d}^4 k$
stands for the integral over loop momenta $k^\mu = (\omega,\mathbf{k})$.

For a static uniform system, where the order parameter $\Delta(x) =
\Delta$ is independent of spacetime, the trace in Eq.~(\ref{Sgeneff})
reduces to
\begin{equation} 
\label{Truni}
\Tr \ln \left[K(p,\Delta) \right] =
\tr \int \mathrm{d}^4 x \int \frac{\mathrm{d}^4 k}{(2
  \pi)^4} \,  \, \ln \left[K(k,\Delta) \right]
\end{equation} 
with the integral over spacetime giving just a volume factor.  The
frequency integral in Eq.~(\ref{312}) can be evaluated in closed form to
give for the one-fermion-loop effective action $S[\Delta, \Delta^*] = -
\int \mathrm{d} t \mathrm{d}^3 x \, \mathcal{V}_{\rm eff}$ with
\begin{equation}  
\label{bcs:closed}
\mathcal{V}_{\rm eff} = -\frac{1}{\lambda} |\Delta|^2 - \int
\frac{\mathrm{d}^3k}{(2 \pi)^3} \left[ E({\bf k}) - \xi({\bf k})
  \right],
\end{equation}
the effective energy density and
\begin{equation}
\label{bcs:Ek}
E({\bf k}) \equiv \sqrt{\xi^2({\bf k}) + |\Delta|^2}
\end{equation}
the single-fermion excitation spectrum of the noninteracting system.
The celebrated BCS gap equation,
\begin{equation}
\label{bcs:gape}
- \frac{1}{\lambda} = \frac{1}{2} \int \frac{\mathrm{d}^3 k}{(2 \pi)^3}
\frac{1}{E({\bf k})},
\end{equation}
follows by minimizing the effective potential with respect to $\Delta$.
The integral appearing here can be evaluated analytically to
give~\cite{Marini:1998}
\begin{equation}
\label{gapeq} 
\frac{1}{k_\mathrm{F} a} = \frac{4}{\pi} \, \frac{I_5(x_o) - x_o
  I_6(x_o)}{\left[x_o I_5(x_o) + I_6(x_o)\right]^{1/3}},
\end{equation} 
where $x_o \equiv \mu/\Delta$, and $I_5$ and $I_6$ are two integrals
which are related to the response functions $\partial^2 \mathcal{V}_{\rm
  eff}/\partial \mu^2$, or $\partial^2 \mathcal{V}_{\rm eff}/\partial
\Delta^2$ and $\partial^2 \mathcal{V}_{\rm eff}/\partial \Delta \partial
\mu$, respectively, and which can be expressed in terms of Legendre
functions \cite{Papenbrock:1999}, see the Appendix.  The momentum
$k_\mathrm{F}$, defined through
\begin{equation} 
n \equiv \frac{1}{3 \pi^2} k_\mathrm{F}^3 ,
\end{equation} 
is used to remove the dimension of the scattering length introduced in
Eq.~(\ref{lar}).  This parameter relates the particle number density $n$
of the interacting Fermi gas to the Fermi momentum of a free Fermi gas
at the same density.  The density $n$ of the interacting system is
obtained from the effective potential (\ref{bcs:closed}) by
differentiating it with (minus) the chemical potential.  The result can
be put in the form~\cite{Marini:1998} 
\begin{equation} 
\label{neq}
  \frac{\Delta}{\epsilon_\mathrm{F}} = \frac{1}{\left[x_o I_5(x_o) +
      I_6(x_o)\right]^{2/3}} ,
\end{equation} 
with $\epsilon_\mathrm{F} \equiv k_\mathrm{F}^2/2m = (3 \pi^2
n)^{2/3}/2m$ the Fermi energy of a free gas.  Equations~(\ref{gapeq})
and (\ref{neq}) are valid in the entire BCS-BEC crossover, with the BCS
limit ($\mu>0, \; a<0$) corresponding to $x_o >> 1$, where 
\begin{equation} 
  I_5(x_o) \approx x_o^{1/2} , \qquad I_6(x_o) \approx \ln ( 8 x_o )/2
  x_o^{1/2} ,
\end{equation} 
and the BEC limit ($\mu<0, \; a>0$) corresponding to $x_o << - 1$, where 
\begin{equation} 
  I_5(x_o) \approx \pi/16 |x_o|^{3/2}, \qquad I_6(x_o) \approx \pi/4
  |x_o|^{1/2} .
\end{equation} 
At unitarity, where a bound state appears and $1/k_\mathrm{F} a$
tends to zero, $I_5 = x_o I_6$ by Eq.~(\ref{gapeq}), so that $x_o =
0.8604\ldots$, $I_5 = 0.8693\ldots, \; I_6 = 1.010\ldots$, and
\begin{equation} 
 \frac{\Delta}{\epsilon_\mathrm{F}} =
 \frac{1}{\left[(1+x_o^2)I_6\right]^{2/3}} = 0.6864\ldots.
\end{equation} 
For the ratio $\xi$ of the chemical potential and the Fermi energy
$\epsilon_\mathrm{F}=k_\mathrm{F}^2/2m$ of the free gas, this gives
the value
\begin{equation}
\label{xi}
\xi \equiv \frac{\mu}{\epsilon_\mathrm{F}} = x_o
\frac{\Delta}{\epsilon_\mathrm{F}} =
\frac{x_o}{\left[(1+x_o^2)I_6\right]^{2/3}} = 0.5906\ldots .
\end{equation}
These mean-field values, which were first obtained numerically in
Ref.~\cite{Engelbrecht:1997}, should be compared with, for
example, the estimates $\Delta/\epsilon_\mathrm{F} = 0.84(4)$ and $\xi =
0.42(1)$ obtained through quantum Monte Carlo simulations of systems of
about 60 particles \cite{Carlson:2005}.

One of the observations of Ref.~\cite{Son:2006} is that all the
leading order terms in the effective theory describing the
Anderson-Bogoliubov mode at unitarity can be determined from the
expression (\ref{bcs:closed}) for the static uniform system.  This will
be demonstrated in the following section by explicit calculation.

To determine the next-to-leading terms, the auxiliary fields $\Delta$
and $\Delta^\dagger$ can no longer be assumed to be constant in the
formal expression (\ref{312}).  It can then at best be evaluated in a
\textit{derivative expansion}.  The method
\cite{Fraser:1984zb,Aitchison:1985pp} we adopt proceeds as follows.
First, the logarithm is expanded in a Taylor series.  Each term in the
series contains powers of the derivative $p^\mu$ operating on
\textit{every} field appearing to the right.  Second, all these
operators are shifted to the left by repeated use of the identity
\begin{equation}
\label{cal:id}
\phi_1(x) p^\mu \phi_2(x) = (p^\mu - \mathrm{i}  \partial^\mu)
\phi_1(x) \phi_2(x),
\end{equation}
where $\phi_1(x)$ and $\phi_2(x)$ are arbitrary fields, and the
derivative $\partial^\mu=(\partial_t,-\nabla)$ acts by convention
\textit{only} on the next field to the right.  Next, each term is
integrated by parts so that all the $p^\mu$'s act to the left where only
a factor $\exp(\mathrm{i} k \cdot x)$ is present and yield a factor of
$k^\mu$.  In this way, each occurrence of the operator $p^\mu$ is replaced with
an integration variable $k^\mu$.  Finally, the exponential function
$\exp(-\mathrm{i} k \cdot x)$ is moved to the left where it is
multiplied with $\exp(\mathrm{i} k \cdot x)$ to give unity.  The
momentum integration can now in principle be carried out to yield an
effective action written as a spacetime integral over a Lagrangian
density, $S = \int \mathrm{d}^4 x \, {\cal L}$.

\section{Leading order}
\label{sec:LO}
To derive the effective action, we write the complex field $\Delta(x)$
in terms of a spacetime-dependent amplitude and phase as
\begin{equation} \label{bcs:London}
\Delta(x) = [\Delta + \sigma(x)] \, {\rm e}^{2 \mathrm{i} \varphi (x)} ,
\end{equation}
with $\Delta$ denoting a real solution of the gap equation
(\ref{bcs:gape}).  The functional measure in Eq.~(\ref{37}) must be
changed accordingly by expressing it in terms of the new fields.
Physically, the phase $\varphi(x)$ of the order parameter describes the
Nambu-Goldstone mode accompanying the spontaneous breakdown of global
U(1) phase symmetry.  The effective action governing these phonons
obtains after integrating out the $\sigma$ field.

The phase can be removed from the order parameter by returning to the
expression (\ref{bcs:36}) for the partition function and decompose the
fermion fields as
\begin{equation} 
\psi_\sigma(x) = {\rm e}^{\mathrm{i} \varphi (x)} \chi_\sigma(x) .
\end{equation} 
Instead of the one-fermion-loop effective action (\ref{312}), one
then obtains
\begin{equation}
\label{SsigmaV}
S[\sigma,\varphi] = \frac{1}{\lambda} \int \mathrm{d}^4 x \, (\Delta +
\sigma)^2 - \mathrm{i} \Tr \ln \left(
\begin{array}{cc} p_{0} - V_0(x) - \xi [{\bf p} + {\bf V}(x)] 
& -[\Delta + \sigma(x)] \\ -[\Delta + \sigma(x)] & p_{0} + V_0 + \xi
  [{\bf p} - {\bf V}(x)]
\end{array} \right),
\end{equation}
where $V_\mu(x) \equiv \partial_\mu \varphi(x)$ formally plays the role
of an Abelian gauge field. In this guise, the Nambu-Goldstone field is
invariably accompanied by at least one derivative.  The resulting
effective theory is thus automatically invariant under global U(1) phase
transformations, under which $\varphi(x)$ is shifted by a constant,
\begin{equation}
\varphi(x) \to \varphi(x) + \mathrm{const.}
\end{equation}  
By construction, the $\sigma$ field appears only in the combination
$\Delta + \sigma$.

The leading order (LO) terms in the effective theory can be obtained by
ignoring derivatives on $\sigma$ and $V_\mu$ so that, after using
Eq.~(\ref{Truni}), $\xi ({\bf k} \pm {\bf V}) = \xi ({\bf k}) + {\bf
  V}^2/2m \pm \mathbf{k} \cdot \mathbf{V}/m$ in Eq.~(\ref{SsigmaV}).  We
explicitly checked that the terms $\pm \mathbf{k} \cdot \mathbf{V}/m$,\
do not contribute in leading order.  The constant field $V_\mu$ thus
appears only in the combination $X \equiv \mu - V$ with
\begin{equation}
\label{Vdef} 
  V \equiv  V_0 + \frac{1}{2m} \mathbf{V}^2 = 
  \partial_t \varphi + \frac{1}{2m} (\nabla \varphi)^2 .
\end{equation} 
By Eq.~(\ref{bs:phiboost}), the Nambu-Goldstone field transforms under a
Galilei boost as
\begin{equation}
\label{cl:vaphiinv}
\varphi(x) \rightarrow \varphi'(x') = \varphi(x) - m  {\bf u}
\cdot {\bf x} + \tfrac{1}{2} m  {\bf u}^2 t ,
\end{equation}
and the two terms at the right of Eq.~(\ref{Vdef}) combine precisely
so that $V$ is invariant. 

Given these observations, the LO terms, i.e., terms without derivatives
on $\sigma$ and $V$, in the one-fermion-loop effective action
$S[\sigma,\varphi] = \int \mathrm{d}^4 x \mathcal{L}_\mathrm{LO} +
\ldots$ governing these fields, must be of the form
\begin{equation} 
\label{lo}
  \mathcal{L}_\mathrm{LO} (\sigma, \varphi) = \sum_{i,j=0}^\infty
  \frac{1}{i! j!}  \pi^{(i,j)} \sigma^i V^j
\end{equation} 
with $\pi^{(i,j)}$ the expansion coefficients, or vertices 
\begin{equation} 
\pi^{(i,j)} = (-1)^{j+1} \frac{\partial^{i+j} }{\partial \Delta^i
  \partial \mu^j} \mathcal{V}_{\rm eff} ,
\end{equation} 
which depend on the parameters $m,\mu,\Delta$ of the theory.  By
dimensional analysis,
\begin{equation} 
\label{coef}
\pi^{(i,j)} = \frac{1}{2^{1/2} \pi^2} m^{3/2} \Delta^{5/2-i-j}
\bar{\pi}^{(i,j)} (x_o)
\end{equation}  
with $x_o = \mu/\Delta$.  The numerical prefactor is pulled out for
later convenience.  The expansion coefficients can be readily obtained from
the expression (\ref{bcs:closed}) for the effective potential.  The
results up to order $i+j=4$ are recorded in the Appendix.  We iterate
that the only terms generated at leading order are those dictated by
symmetry and are precisely the once included in Eq.~(\ref{lo}).
Returning to the original expression (\ref{SsigmaV}), we explicitly
checked that terms spoiling any of the symmetries drop out.

To the order $i+j=2$, i.e., in the Gaussian approximation, the $\sigma$
field can be integrated out by substituting the corresponding
Euler-Lagrange equation for this field,
\begin{equation} 
\label{EL}
\sigma = - \frac{\pi^{(1,1)}}{\pi^{(0,2)}} V = - \frac{I_6}{I_5} V
\end{equation} 
back into the Lagrangian (\ref{lo}) with $i+j\leq2$.  This is tantamount
to approximating the integral over $\sigma$ by the saddle point.  It
gives as effective theory governing solely the Anderson-Bogoliubov mode
at LO 
\begin{equation} 
\label{LO2}
  \mathcal{L}_\mathrm{LO}(\varphi) = \pi^{(0,1)} V - \frac{1}{2}
  \frac{\left(\pi^{(1,1)}\right)^2 - \pi^{(0,2)}
    \pi^{(2,0)}}{\pi^{(2,0)}} V^2 .
\end{equation} 
From it, the speed of propagation $c$ of this gapless mode can be read
off as
\begin{equation} 
c^2 =  \frac{1}{m} \frac{\pi^{(0,1)}
  \pi^{(2,0)}}{\left(\pi^{(1,1)}\right)^2 - \pi^{(0,2)} \pi^{(2,0)}} .
\end{equation} 
Substituting the explicit expressions (\ref{clo}) for the coefficients
$\pi^{(i,j)}$,  we reproduce the result due to Marini, Pistolesi, and
Strinati~\cite{Marini:1998}  
\begin{equation} 
\label{c}
c^2 = \frac{2}{3} \frac{\mu}{m} \frac{I_5(x_o I_5 + I_6)}{x_o(I_5^2 +
  I_6^2)} .
\end{equation} 
In the weak-coupling BCS limit, obtained by letting $x_o \to \infty$, as
well as in the strong-coupling unitary limit, where $I_5 = x_o I_6$ with
$x_o = 0.8604\ldots$, it reduces to the same form
\begin{equation}
\label{c2}
  c^2 = \frac{2}{3} \frac{\mu}{m} .
\end{equation} 
Both limits are in the regime where the chemical potential is
positive.  At unitarity,
\begin{equation}
\label{c2u} 
\frac{c^2}{v_\mathrm{F}^2} = \frac{1}{3} \xi = 0.1968\ldots, 
\end{equation} 
with $v_\mathrm{F} \equiv k_\mathrm{F}/m$ the Fermi velocity of the free
Fermi gas at the same density as the unitary gas, and $\xi$ the
dimensionless parameter (\ref{xi}).

We next turn to LO terms of higher powers in the fields.  In the BCS
limit $x_o \to \infty$, where at leading order in energy and momentum,
the $\sigma$ field decouples from $V$ and can be ignored, the effective
theory (\ref{lo}) up to quartic order in $V$ reduces to
\begin{equation} 
\label{loBCS} 
  \mathcal{L}_\mathrm{LO}(\varphi) = \frac{2^{5/2}}{15 \pi^2} m^{3/2}
  \mu^{5/2} \left(1 - \frac{5}{2} \hat{V} + \frac{15}{8} \hat{V}^2 -
  \frac{5}{16} \hat{V}^3 - \frac{5}{128} \hat{V}^4 + \ldots \right) ,
\end{equation} 
where we added the free Fermi gas contribution and introduced the
abbreviation $\hat{V} \equiv V/\mu$.  In this limit, $\mu \to
\epsilon_\mathrm{F}$.  The terms in Eq.~(\ref{loBCS}) form precisely the
first in the Taylor series expansion of the full LO expression,
\begin{equation} 
\label{sowi}
\mathcal{L}_\mathrm{LO}(\varphi) = c_0 m^{3/2} X^{5/2}, \qquad X \equiv
\mu - V = \mu - \partial_t \varphi - \frac{1}{2m} (\nabla \varphi)^2 ,
\end{equation} 
proposed by Son and Wingate \cite{Son:2006} with the coefficient
\begin{equation} 
\label{c_0bcs}
c_0 = \frac{2^{5/2}}{15 \pi^2}.
\end{equation} 
The expression (\ref{sowi}) sums up all terms where each
Nambu-Goldstone field is accompanied by exactly one (space or time)
derivative.  For a static uniform system, where $V$ is zero, the right
side of Eq.~(\ref{sowi}) reduces to the zero-temperature pressure
expressed as a function of $\mu$, $P(\mu)$.

As an aside, a trapping or other external potential $U(x)$ can be
readily included in the formalism by replacing $\mu$ with $\mu - U(x)$
in $X$.

To obtain the effective theory governing the field $V$ alone in the
entire BCS-BEC region, the $\sigma$ field must be integrated out from
the complete LO expression (\ref{lo}).  Since it contains arbitrary
powers of $\sigma$, this is in general impossible.  At unitary, however,
the coefficients assume values that make this at least approximately
possible, as we now demonstrate.  The Lagrangian up to quartic order in
$\sigma$ and $V$, with the vertices given in the Appendix, takes the
following form in this limit:
\begin{eqnarray}  
\label{LsigmaV}
\mathcal{L}_\mathrm{LO} (\sigma, \varphi) = \frac{2^{1/2}}{\pi^2}
m^{3/2} \mu^{5/2} \frac{I_6}{x_o^{3/2}} &\biggl[& -\frac{2}{3} (1 +
  x_o^2) \hat{V} + \frac{1}{2} x_o^2 \hat{V}^2 - \frac{1}{12} x_o^2
  \hat{V}^3 - \frac{1}{96} \frac{x_o^4}{1+x_o^2} \hat{V}^4 \nonumber
  \\ &&- x_o \hat{\sigma} \hat{V} - \frac{1}{2} x_o^2 \hat{\sigma}^2 -
  \frac{1}{6} x_o^3 \hat{\sigma}^3 + \frac{1}{96} \frac{x_o^4(3 + 4
    x_o^2)}{1+x_o^2} \hat{\sigma}^4 \nonumber \\ && - \frac{1}{4} x_o^2
  \hat{\sigma}^2 \hat{V} + \frac{1}{24} \frac{x_o^3}{1+x_o^2}
  \hat{\sigma} \hat{V}^3 + \frac{1}{16} \frac{x_o^4}{1+x_o^2}
  \hat{\sigma}^2 \hat{V}^2 \nonumber \\ && + \frac{1}{24}
  \frac{x_o^3(1+2 x_o^2)}{1+x_o^2} \hat{\sigma}^3 \hat{V} + \ldots
  \biggr] .
\end{eqnarray} 
Note that the vertex $\pi^{(1,2)}$ of the $\sigma V^2$ term vanishes
in this limit.  Moreover, the coefficients are such that the
Euler-Lagrange equation (\ref{EL}), which assumes the form $\sigma = -
V/x_o$ in the unitary limit, obtained in the Gaussian approximation
remains unchanged after including the additional terms in
Eq.~(\ref{LsigmaV}).  Put differently, the saddle point of the
\textit{nonlinear} theory~(\ref{LsigmaV}) remains locked at $\sigma = -
V/x_o$ in the unitary limit.  Approximating the integral over $\sigma$
by this saddle point, we obtain as effective Lagrangian governing just
$\varphi$
\begin{equation} 
   \mathcal{L}_\mathrm{LO} (\varphi) = \frac{2^{5/2}}{15 \pi^2} m^{3/2}
  \mu^{5/2} \frac{(1+x_o^2)I_6}{x_o^{3/2}} \left(1 - \frac{5}{2} \hat{V}
  + \frac{15}{8} \hat{V}^2 - \frac{5}{16} \hat{V}^3 - \frac{5}{128}
  \hat{V}^4 + \ldots \right) ,
\end{equation} 
where we included the term for a static uniform unitary Fermi gas.  The
various contributions combine to exactly generate the first terms in the
Taylor series expansion of the predicted form (\ref{sowi}) with
\begin{equation}
\label{c_0uni} 
  c_0 = \frac{2^{5/2}}{15 \pi^2} \frac{1}{\xi^{3/2}} , \qquad
  \frac{1}{\xi^{3/2}} \equiv \frac{(1+x_o^2)I_6}{x_o^{3/2}} =
  2.203\ldots ,
\end{equation} 
so that $c_0 = 0.0841\ldots$.  The dimensionless parameter $\xi$, which
was introduced in Eq.~(\ref{xi}) as the ratio of the chemical potential
and $\epsilon_\mathrm{F}$, gives here the ratio of the ground-state
energy per particle $\epsilon$ of the unitary gas and that of a free Fermi
gas, or equivalently, a neutral BCS superconductor at the same density,
\begin{equation} 
\epsilon  \equiv \xi \frac{3}{10} \frac{k_\mathrm{F}^2}{m} ,
\end{equation}
as follows from taking the Legendre transform of $P(\mu)$ and using that
$n = \partial P(\mu)/\partial \mu$ \cite{Son:2006}.  We emphasize that
in determining the coefficients of the LO terms only the effective
potential (\ref{bcs:closed}) describing a static uniform system is used.
This validates the symmetry argument by Son and Wingate \cite{Son:2006}
that the LO effective theory (\ref{sowi}) to all orders in the
Nambu-Goldstone field can be obtained by simply replacing the chemical
potential $\mu$ with the Galilei-invariant combination $X$ in the
pressure $P(\mu)$ of the static uniform system.  It implies that
the complete interaction between phonons at leading order in wave vector
and frequency is determined by $P(\mu)$.

The strong-coupling unitary limit is special as for no other point in
the BCS-BEC crossover, the saddle point (\ref{EL}), obtained in the
Gaussian approximation, constitutes a solution to the Euler-Lagrange
equation when additional terms in the Lagrangian (\ref{lo}) are
included.

\section{1D}
\label{sec:1D}

Although the coefficients of the LO effective Lagrangian in three space
dimensions (3D) can only be determined approximately in the
strong-coupling unitary limit, the form of the theory is precisely as
predicted by Son and Wingate \cite{Son:2006}.  To provide further
support for this prediction, we in this section consider the pairing
theory in one space dimension (1D) for which exact results are
available.

The 1D system of $N$ spin-$\frac{1}{2}$ fermions with attractive
delta-function interactions is described exactly by the Gaudin integral
equations \cite{Gaudin:1967}.  Let, as in the three-dimensional case,
$\lambda (< 0)$ denote the coupling constant.  It is related to the
1D scattering length through
\begin{equation} 
\frac{1}{\lambda} = - \frac{1}{2} m a ,
\end{equation} 
as follows from Eq.~(\ref{lar}) with $D=1$.  Whereas the 3D coupling
constant is directly proportional to the scattering length, its 1D
counterpart is inversely proportional to $a$, and $a \to + \infty$ in
the limit $\lambda \to 0^-$.  This divergence of the scattering length
in the zero-coupling limit arises because an attractive delta-function
potential possesses a two-particle bound state however small the
attraction.  In other words, this limit marks the threshold of a bound
state in vacuum, where the scattering length diverges and changes from
negative for repulsive interactions ($\lambda >0$) to positive for
attractive interactions ($\lambda <0$).

In the weak-coupling BCS limit $\lambda \to 0^-$, the Gaudin integral
equations yield for the ground-state energy per particle $\epsilon$,
chemical potential $\mu$, and sound velocity $c$, the free Fermi gas
expressions
\begin{equation}
\label{free} 
\epsilon = \frac{1}{6} \frac{k_\mathrm{F}^2}{m}, \quad \mu = \epsilon_\mathrm{F},
\quad c^2 = 2 \frac{\mu}{m} = v_\mathrm{F}^2 ,
\end{equation} 
as expected.  In the strong-coupling limit $\lambda \to -\infty$, where
$a \to 0^+$ in 1D, the fermions form tightly bound pairs with binding
energy $\epsilon_a = 1/m a^2$, as in 3D. The Gaudin integral equations
give in this limit \cite{Astrakharchik:2004}
\begin{equation} 
\label{unitary_1D}
\epsilon_\mathrm{eff} = \frac{1}{24} \frac{k_\mathrm{F}^2}{m}, \quad
\mu_\mathrm{eff} = \frac{1}{8} \frac{k_\mathrm{F}^2}{m}, \quad c^2 = 2
\frac{\mu_\mathrm{eff}}{m} = \frac{1}{4} \frac{k_\mathrm{F}^2}{m^2} ,
\end{equation} 
where we removed the (diverging) binding energy from the ground-state
energy and the chemical potential by introducing $\epsilon_\mathrm{eff}
\equiv \frac{1}{2} \epsilon_a + \epsilon$ and $\mu_\mathrm{eff} \equiv
\frac{1}{2} \epsilon_a + \mu$.  The Fermi momentum $k_\mathrm{F}$ is now
defined through
\begin{equation} 
n \equiv \frac{2}{\pi} k_\mathrm{F}.
\end{equation} 
Note that as in 3D, the unitary limit is in the region with positive
(effective) chemical potential.  On comparison with the weak-coupling
results (\ref{free}), it follows that
\begin{equation} 
\xi \equiv \frac{\mu_\mathrm{eff}}{\epsilon_\mathrm{F}} = \frac{1}{4}
\end{equation} 
exactly.  As in 3D, $\xi$ also equals the ratio of the (effective)
ground-state energy per particle $\epsilon$ of the unitary gas and that
of a free Fermi gas at the same density.  The exact 1D counterpart of
Eq.~(\ref{c2u}) reads
\begin{equation}
\label{c2u1} 
\frac{c^2}{v_\mathrm{F}^2} = \xi = \frac{1}{4}  ,
\end{equation} 
whereas the full LO Lagrangian density is given by
\begin{equation} 
\label{sowi_1D}
\mathcal{L}_\mathrm{LO}(\varphi) = c_0 m^{1/2} X^{3/2},
\end{equation} 
with 
\begin{equation}
\label{c_0uni_1D} 
  c_0 = \frac{2^{5/2}}{3 \pi} \frac{1}{\xi^{1/2}} = \frac{2^{7/2}}{3 \pi} 
\end{equation} 
exactly in the unitary limit (and $c_0= 2^{5/2}/3 \pi$ in the
weak-coupling limit).  

The physical interpretation of the 1D unitary limit follows from
comparison with the related problem of repulsively interacting
bosons.  The 1D bosonic system with a delta-function potential is described
exactly by the Lieb-Liniger integral equations \cite{Lieb:1963a}.  In
the infinite-coupling limit, these equations coincide with the Gaudin
integral equations at unitary, provided one identifies the boson mass
$m_\mathrm{b}$ with twice the fermion mass, $m_\mathrm{b}=2 m$, and the
boson number density $n_\mathrm{b}$ with half the fermion number
density, $n_\mathrm{b}=n/2$ \cite{Astrakharchik:2004}.  That is, the 1D
unitary limit coincides with the BEC limit.

The two sets of integral equations can be mapped onto each other not
only in the infinite-coupling limit, but also for finite coupling.  With
the above identifications, both sets become similar with one important
distinction that the signs of the coupling constants differ, as already
pointed out by Gaudin \cite{Gaudin:1967}.  Specifically, the Gaudin
integral equations can be obtained from the Lieb-Liniger integral
equations by setting $m_\mathrm{b}=2 m$ and $n_\mathrm{b}=n/2$, and by
replacing the bosonic coupling constant $\lambda_\mathrm{b}(>0)$ with
$-2 \lambda(<0)$ \cite{Iida:2005}, showing that the interaction
between pairs is \emph{attractive}.  Whereas bosons must interact
repulsively to guarantee stability, pairs of spin-$\frac{1}{2}$
particles can have attractive interactions, for the Pauli exclusion
principle forbids two such pairs to form a four-fermion bound state.
The fact that $c^2$ is positive in the unitary limit implies that the
compressibility is also positive and that the system is mechanically
stable even in this limit of infinite attraction between pairs.

\section{Next-to-leading order}
\label{sec:NLO}
We next turn to the next-to-leading order (NLO) terms in the effective
Lagrangian.  These terms involve derivatives of the $\sigma$ field and
$V$, or equivalently, $X = \mu - V$.  To obtain the first NLO terms it
suffice to consider quadratic terms in the fields $\sigma$ and $\varphi$
up to fourth order in derivatives.  The possible independent terms of
this form are given in 3D by
\begin{eqnarray} 
\label{qua}
 \mathcal{L}_\mathrm{NLO}^{(2)}(\sigma, \varphi) = \frac{1}{2^{1/2} \, 6
   \pi^2} & \biggl[ & b_1 \left(\frac{m}{\Delta}\right)^{3/2}
   (\partial_t \sigma)^2 + b_2 \left(\frac{m}{\Delta}\right)^{1/2}
   (\nabla \sigma)^2 \nonumber \\ && + c_1
   \left(\frac{m}{\Delta}\right)^{1/2} (\nabla \partial_t \varphi)^2 +
   c_2 \left(\frac{m}{\Delta}\right)^{3/2} (\partial^2_t \varphi)^2 +
   c_3 \left(\frac{\Delta}{m}\right)^{1/2} (\nabla^2 \varphi)^2 + c_4
   \left(\frac{m}{\Delta}\right)^{1/2} \partial_t^2 \varphi \nabla^2
   \varphi \nonumber \\ && + d_1 \left(\frac{m}{\Delta}\right)^{1/2}
   \nabla \sigma \cdot \nabla \partial_t \varphi + d_2
   \left(\frac{m}{\Delta}\right)^{3/2} \partial_t \sigma \partial_t^2
   \varphi + d_4 \left(\frac{m}{\Delta}\right)^{1/2} \partial_t \sigma
   \nabla^2 \varphi \biggr],
\end{eqnarray}
where the powers of $m/\Delta$ follow from dimensional analysis.  In
writing this general expression, with arbitrary coefficients $b_i, c_i$,
and $d_i$, we also used that the theory is invariant under time
reversal, under which $t \to -t$ and $\varphi \to - \varphi$.  Our
choice of coefficients is such that at unitarity, the terms with
coefficients $d_1,d_2$ and $d_4$ combine with those with coefficients
$c_1,c_2$ and $c_4$, respectively after the $\sigma$ field has been
integrated out.  In addition to the $(\nabla^2 \varphi)^2$ term, there
exists a second term quartic in derivatives, \textit{viz.}  $\partial_i
\partial_j \varphi \partial_i \partial_j \varphi$.  In the quadratic
approximation we are working, both differ by a total derivative and
cannot be uniquely identified.  However, the ratio of the two terms can
be determined by the derivative expansion method we use and comes out to
be two, leaving two possible combinations $(\nabla^2 \varphi)^2 + 2
(\partial_i \partial_j \varphi)^2$ or $2(\nabla^2 \varphi)^2 +
(\partial_i \partial_j \varphi)^2$.  Through the study of two
three-point correlation functions, it was shown in
Ref.~\cite{Manes:2009} that the former combination is in fact
realized.  This combination is also favored by symmetry, for
\begin{equation} 
\label{perm}
 (\nabla^2 \varphi)^2 + 2 (\partial_i \partial_j \varphi)^2 =
  (\delta_{i j} \delta_{kl} + \delta_{i k} \delta_{jl} + \delta_{i l}
  \delta_{kj}) \partial_i \partial_j \varphi \partial_k \partial_l
  \varphi .
\end{equation} 

We have computed the coefficients appearing in the
Lagrangian (\ref{qua}) for the entire BCS-BEC crossover by applying the
derivative expansion method to the formal expression (\ref{SsigmaV}).
The results are collected in the Appendix.  Although the terms $(\nabla
\partial_t \varphi)^2$ (with coefficient $c_1$) and $\partial_t^2 \varphi
\nabla^2 \varphi$ (with coefficient $c_4$) become identical after partial
integration, they can be separately identified, for they originate from
different parts in the the effective action (\ref{SsigmaV}).
Specifically, the term $(\nabla
\partial_t \varphi)^2$ arises as $(\nabla V_0)^2$ and can be calculated
by setting $\mathbf{V}$ to zero in Eq.~(\ref{SsigmaV}) and ignoring
further time derivatives, while the term $\partial_t^2 \varphi \nabla^2
\varphi$ arises as $\partial_t V \nabla \cdot \mathbf{V}$.

We first consider the BCS limit, where the $\sigma$ field decouples in
first approximation.  The general expression (\ref{qua}) reduces in this
limit, obtained by letting $x_o \to \infty$, to the known result (in
conventional notation) \cite{Andrianov:1976,Popov:1976,effBCS}
\begin{equation}
\label{bcs_qua} 
\mathcal{L}_\mathrm{NLO}^{(2)}(\varphi) = \frac{1}{6}
\frac{\nu(0)}{\Delta^2} \left[(\partial_t^2 \varphi)^2 + \frac{1}{5}
  v_\mathrm{F}^4 (\nabla^2 \varphi)^2 - \frac{2}{3} v_\mathrm{F}^2
  \partial_t^2 \varphi \nabla^2 \varphi \right] ,
\end{equation} 
with $\nu(0) \equiv m k_\mathrm{F}/2 \pi^2$ the density of states at the
Fermi surface.  Note that in this limit, the coefficient $c_1$ can be
ignored in comparison to the coefficient $c_4$.  These NLO terms
together with the quadratic LO terms in Eq.~(\ref{LO2}) give rise to the
phonon spectrum \cite{Andrianov:1976,Popov:1976}
\begin{equation} 
\label{gammaBCS} 
\omega^2(\mathbf{k}) = \frac{1}{3} v_\mathrm{F}^2 \mathbf{k}^2
\left( 1 - \frac{2}{45} \frac{v_\mathrm{F}^2}{\Delta^2} \mathbf{k}^2
\right) = \frac{2}{3} \frac{\mu}{m} \mathbf{k}^2 \left( 1
- \frac{4}{45} \frac{x_o^2}{m \mu} \mathbf{k}^2 \right) .
\end{equation} 
The quadratic terms (\ref{bcs_qua}) can be put into Galilei-invariant
form as
\begin{equation} 
\label{NLObcs}
  \mathcal{L}_\mathrm{NLO}(\varphi) = \frac{5}{16} c_0 x_o^2 \left\{
  \frac{m^{3/2}}{X^{3/2}} (D_t X)^2 + \frac{4}{15}
  \frac{X^{1/2}}{m^{1/2}} \left[(\nabla^2 \varphi)^2 + 2 (\partial_i
    \partial_j \varphi)^2 \right] + \frac{4}{3} \frac{m^{1/2}}{X^{1/2}}
  D_t X \nabla^2 \varphi \right\},
\end{equation} 
with $c_0$ given in Eq.~(\ref{c_0bcs}) and $D_t$ the 
material derivative,
\begin{equation} 
D_t \equiv \partial_t + \frac{1}{m} \nabla \varphi . \nabla \; ,
\end{equation} 
which by the transformations (\ref{bs:boostp}) and (\ref{cl:vaphiinv})
is invariant under Galilei boosts.  We note that the coefficients of the
NLO terms can be uniquely identified from the quadratic approximation
(\ref{bcs_qua}) to this Lagrangian.  Indeed, as prescribed by the
derivative expansion method \cite{Fraser:1984zb}, each factor of $\mu$
is to be replaced with $X$ in Eq.~(\ref{NLObcs}).  And each
occurrence of $\partial_t^2 \varphi$ is to be replaced with the
Galilei-invariant form $-D_t X$.
   
For completeness, we mention that the Lagrangian governing the $\sigma$
field in the BCS limit reads to this order
\begin{equation}
 \mathcal{L}(\sigma) = \frac{2^{1/2}}{\pi^2} m^{3/2} \mu^{5/2}
\frac{1}{x_o^2} \left\{ \frac{1}{24} \frac{1}{\Delta^2}\left[
  (\partial_t \bar{\sigma})^2 - \frac{1}{3} v_\mathrm{F}^2 (\nabla
  \bar{\sigma})^2 \right] - \frac{1}{2} \bar{\sigma}^2 - \frac{1}{3!}
\bar{\sigma}^3 + \frac{1}{4!}  \bar{\sigma}^4 + \ldots \right\},
\end{equation} 
where $\bar{\sigma} \equiv \sigma/\Delta$, and vanishes in the limit
$x_o \to \infty$.

We continue by studying the strong-coupling unitary limit obtained by
setting $I_5 = x_o I_6$ with $x_o = 0.8604\ldots$.  The general
expression (\ref{qua}) assumes in this limit the explicit form:
\begin{eqnarray} 
 \mathcal{L}_\mathrm{NLO}^{(2)}(\sigma, \varphi) = \frac{1}{2^{1/2} \, 3
   \pi^2} x_o^{1/2} I_6 &\biggl[&\frac{1}{4}
   \left(\frac{m}{\mu}\right)^{3/2} x_o^2 (\partial_t \sigma)^2 -
   \frac{1}{24} \left(\frac{m}{\mu}\right)^{1/2} (7 + 4 x_o^2) (\nabla
   \sigma)^2 \nonumber \\ && - \frac{1}{8}
   \left(\frac{m}{\mu}\right)^{1/2} (\nabla \partial_t \varphi)^2 +
   \frac{1}{2} x_o^2 \left(\frac{m}{\mu}\right)^{3/2} (\partial^2_t
   \varphi)^2 \nonumber \\ && + \frac{2}{5}
   (1+x_o^2)\left(\frac{\mu}{m}\right)^{1/2} (\nabla^2 \varphi)^2 -
   \frac{1}{6} (4 x_o^2 + 1)\left(\frac{m}{\mu}\right)^{1/2}
   \partial_t^2 \varphi \nabla^2 \varphi \nonumber \\ && - \frac{1}{4}
   \left(\frac{m}{\mu}\right)^{3/2} x_o^2 \partial_t \sigma \partial_t^2
   \varphi + \frac{1}{2} \left(\frac{m}{\mu}\right)^{1/2} x_o \partial_t
   \sigma \nabla^2 \varphi \biggr] .
\end{eqnarray} 
Note that the vertex $d_1$ of the $\nabla \sigma \cdot \nabla \partial_t
\varphi$ term vanishes in this limit.  Approximating the integral over
$\sigma$ by the LO saddle point (\ref{EL}), which is consistent to the
order we are working, we obtain an effective theory of precisely the form
(\ref{bcs_qua}) with an additional $(\nabla \partial_t \varphi)^2$ term
included.  As in the BCS limit, these quadratic terms fourth order in
derivatives can be put into Galilei-invariant form as
\begin{equation} 
\label{NLOuni}
  \mathcal{L}_\mathrm{NLO}(\varphi) =\frac{5}{16} c_0 x_o^2
  \left\{-\frac{7}{12} \frac{1}{x_o^2} \frac{m^{1/2}}{X^{1/2}} (\nabla
  X)^2 + \frac{m^{3/2}}{X^{3/2}} (D_t X)^2 + \frac{4}{15}
  \frac{X^{1/2}}{m^{1/2}} \left[(\nabla^2 \varphi)^2 + 2 (\partial_i
    \partial_j \varphi)^2 \right] + \frac{4}{3} \frac{m^{1/2}}{X^{1/2}}
  D_t X \nabla^2 \varphi \right\},
\end{equation} 
with $c_0$ now given by Eq.~(\ref{c_0uni}).  Apart from overall
normalization and the first term, which is suppressed in the BCS limit,
these NLO terms are exactly as found in the BCS limit.  As in that
limit, the form and coefficients of the NLO terms uniquely follow from
the quadratic approximation to this Lagrangian.  In addition to the
replacements already used in the BCS limit, $(\nabla \partial_t
\varphi)^2$ is replaced with the Galilei-invariant form $(\nabla X)^2$
in Eq.~(\ref{NLOuni}).  The second and last terms in that expression,
both involving $D_t X$, were not considered by Son and
Wingate~\cite{Son:2006} as their approach is limited to only gradients
of $X$.  Also the fourth term was omitted.  With $r_i$ ($i=1,2,3,4,5$)
denoting the coefficients of the NLO terms, so that
\begin{equation} 
\label{ri}
  r_i = \frac{5}{16} c_0 x_o^2 \left(-\frac{7}{12} \frac{1}{x_o^2}, 1,
  \frac{4}{15}, \frac{8}{15}, \frac{4}{3} \right),
\end{equation} 
the spectrum of the gapless Anderson-Bogoliubov mode that follows when
including the NLO terms (\ref{NLOuni}) reads
\begin{subequations}
\begin{eqnarray} 
\label{gammauni_1}
\omega^2 (\mathbf{k}) &=& \frac{2}{3} \frac{\mu}{m} \mathbf{k}^2 \left\{
  1 - \frac{4}{45 c_0} \left[ 6 r_1 + 4 r_2 + 9 (r_3 +  r_4) - 6 r_5
  \right]\frac{1}{m \mu} \mathbf{k}^2 \right\} \\ &=& \frac{2}{3}
\frac{\mu}{m} \mathbf{k}^2 \left( 1 + \frac{35 - 32 x_o^2}{360}
\frac{1}{m \mu} \mathbf{k}^2 \right) .
\label{gammauni_2}
\end{eqnarray} 
\end{subequations}
It reduces to the BCS expression (\ref{gammaBCS}) in the limit $x_o \to
\infty$.  In contrast to the BCS limit, the coefficient of the
correction term is positive for the value $x_o = 0.8604\ldots$ obtained
in the saddle-point approximation.  If it remains positive beyond this
approximation, a low-energy phonon in a unitary Fermi gas can decay
into two phonons.

The Son-Wingate result for the spectrum corresponds to setting
$r_2=r_5=0$ and also $r_4=0$ in Eq.~(\ref{gammauni_1}), which leads to
an incorrect expression for the spectrum.  The correct result
(\ref{gammauni_2}) was derived from a NLO Lagrangian of a form proposed
by Son and Wingate, i.e., one without time derivatives in 
Ref.~\cite{Manes:2009}.  This was achieved by eliminating the time
derivatives in the NLO Lagrangian through the use of the leading-order
field equations.  This reduction leads to the following changes in the
coefficients (\ref{ri}):
\begin{equation} 
r_3 = \frac{1}{12} c_0 x_o^2 \to r_3' = - \frac{1}{18} c_0 x_o^2 ,
\quad r_4 = 2 r_3 \to r_4' = -3 r_3'
\end{equation} 
which, when substituted in Eq.~(\ref{gammauni_1}) with $r_2=r_5=0$,
yields the correct spectrum (\ref{gammauni_2}).  The flip side of this
reduction is that the static response functions come out incorrectly,
for the coefficients of the static terms in the NLO Lagrangian now also
include dynamic effects.  The relation $r_4' = -3 r_3'$ in the reduced
NLO Lagrangian was argued in Ref.~\cite{Manes:2009} to be a consequence
of conformal invariance at unitarity. 

\section{Discussion}
Its (relatively) simple form and the very fact that the effective action
of a unitary Fermi gas up to next-to-leading order can be derived from
the microscopic theory analytically underscores the special status of
the unitary limit in the BCS-BEC crossover.  Although the coefficients
could only be computed approximately, using an (extended) saddle point,
it is remarkable that the effective field program, which involves
integrating out the fermionic degrees of freedom as well as the $\sigma$
field, can be carried out consistently up to the orders considered,
featuring higher-order terms such as $(\nabla \varphi)^8$ and
$(\partial_t^2 \varphi)^2$.  Surprisingly, the effective actions
obtained in the weak-coupling BCS and the strong-coupling unitary limits
are proportional, save for a next-to-leading order term which is
suppressed in the BCS limit.

\acknowledgments{The author wishes to thank W.~Janke and the other
  members of the Institute of Theoretical Physics, University of Leipzig
  for their kind hospitality and a visiting professorship.  The work at
  the Departamento de F\'{i}sica, Universidade Federal de Pernambuco,
  Recife is financially supported by CAPES, Brazil through a visiting
  professor scholarship.  The author is indebted to G.~L.~Vasconcelos
  for warm hospitality at the UFPE.}

\appendix

\section{Integrals} \label{app:a}
All integrals encountered in this study can be expressed as linear
combinations of two basic integrals introduced by Marini, Pistolesi, and
Strinati~\cite{Marini:1998} 
\begin{equation} 
I_5(x_o) \equiv \int_0^\infty \mathrm{d} x \frac{x^2}{E_x^3}, \qquad
I_6(x_o) \equiv \int_0^\infty \mathrm{d} x \frac{x^2 \xi_x}{E_x^3},
\end{equation} 
by integrating by parts and simple algebraic manipulations.  Here,
$x,\xi_x$, and $E_x$ denote the dimensionless variables
\begin{equation} 
  x^2 \equiv \frac{k^2/2m}{\Delta}, \quad \xi_x \equiv
  \frac{\xi}{\Delta} = x^2 - x_o, \quad x_o \equiv \frac{\mu}{\Delta},
  \quad E_x \equiv \frac{E}{\Delta} = \sqrt{\xi_x^2 +1} .
\end{equation} 
These integrals can be expressed in terms of the complete elliptic
integrals of the first and second kind, as was done in
Ref.~\cite{Marini:1998} or in terms of Legendre functions
$P_\alpha$ \cite{Papenbrock:1999} as ($\gamma_o \equiv
x_o/\sqrt{1+x_o^2}$)
\begin{eqnarray} 
  I_5(x_o) &=& \frac{\pi}{4} \frac{1}{(1+x_o^2)^{3/4}} \left[ (1 - 3
    x_o^2) P_{1/2}\left( - \gamma_o \right) - 3 x_o (1+x_o^2)^{1/2}
    P_{3/2}\left( - \gamma_o \right) \right] \nonumber \\ I_6(x_o) &=& -
  \frac{\pi}{4} \frac{1}{(1+x_o^2)^{3/4}} \left[ 4 x_o P_{1/2}\left(-
    \gamma_o \right) + 3 (1+x_o^2)^{1/2} P_{3/2}\left(- \gamma_o \right)
    \right] ,
\end{eqnarray} 
so that, for example,
\begin{equation} 
I_5(x_o) - x_o I_6 (x_o) = \frac{\pi}{4} (1+x_o^2)^{1/4} P_{1/2}\left(-
    \gamma_o \right) .
\end{equation} 

With the definition 
\begin{equation}
I_{k,l,m} \equiv \int_0^\infty \mathrm{d} x \frac{x^k \xi_x^l}{E_x^m},
\end{equation} 
so that $I_5 = I_{2,0,3}$ and $I_6 = I_{2,1,3}$, one readily verifies the
relations
\begin{subequations}
\begin{eqnarray}
I_{0,0,1} &=& 2 I_6 \\ I_{0,0,3} &=& \frac{x_o I_5 + I_6}{1+x_o^2}
\\ I_{0,0,5} &=& \frac{I_{4,0,5} - 2 I_{2,1,5} + I_{0,0,3}}{1+x_o^2}
\\ I_{0,1,3} &=& \frac{I_5 - x_o I_6}{1+x_o^2} \\ I_{0,2,5} &=& I_{0,0,3} -
I_{0,0,5} \\ I_{0,3,5} &=& I_{2,2,5} - x_o I_{0,2,5} \\ I_{2,0,5} &=&
\frac{1}{6} x_o \frac{x_o I_5 + I_6}{1+x_o^2} + \frac{1}{2} I_5
\\ I_{2,0,7} &=& I_{2,4,7} + 2 I_{2,0,5} - I_5 \\ I_{2,1,5} &=& \frac{1}{6} I_{0,0,3}
\\ I_{2,1,7} &=& \frac{1}{10} I_{0,0,5} \\ I_{2,2,5} &=& \frac{1}{3} I_5 +
\frac{1}{6} I_{0,1,3} \\ I_{2,4,7} &=& \frac{1}{10}( I_{0,3,5} + 6
  I_{2,2,5}) \\ I_{4,0,5} &=& (1+x_o^2) I_{2,1,5} + \frac{1}{2} x_o I_5 .
\end{eqnarray} 
\end{subequations}
These integrals are all a function of $x_o$ alone.

With the help of these integrals, the coefficients $\pi^{(i,j)}$ of the
LO effective theory (\ref{lo}), introduced in Eq.~(\ref{coef}), can be
readily evaluated in closed form, with the results up to $i+j=4$
\begin{subequations}
\label{clo}
\begin{eqnarray} 
\bar{\pi}^{(0,1)} &=& - \frac{4}{3}   (x_o
  I_5 + I_6) \\
\bar{\pi}^{(0,2)} &=&  2
 I_5  \\
\bar{\pi}^{(0,3)} &=& - 
\frac{x_o I_5 + I_6}{1+x_o^2}  \\
\bar{\pi}^{(0,4)} &=& - \frac{1}{2}  
\frac{(-3 + x^2_o) I_5 + 4 x_o I_6}{(1+x_o^2)^2}  \\
\bar{\pi}^{(2,0)} &=& - 2
 I_5 \\
\bar{\pi}^{(3,0)} &=& - 
\frac{ (3 + 2 x_o^2 )I_5 - x_o I_6}{1+x_o^2} \\
\bar{\pi}^{(4,0)} &=&  \frac{1}{2}
  \frac{(3 + 3 x_o^2 + 4 x_o^4 )I_5 +
    4 x_o^3 I_6}{(1+x_o^2)^2} \\
\bar{\pi}^{(1,1)} &=& - 2
 I_6  \\
\bar{\pi}^{(1,2)} &=& 
\frac{I_5 - x_o I_6}{1+x_o^2}  \\
\bar{\pi}^{(1,3)} &=&  \frac{1}{2}  
\frac{4x_o I_5 + (1-3x_o^2) I_6}{(1+x_o^2)^2}  \\
\bar{\pi}^{(2,1)} &=& 
 \frac{x_o I_5 - (1+2 x_o^2)
    I_6}{1+x_o^2} \\ \pi^{(2,2)} &=& - \frac{1}{2} 
 \frac{(1 -3 x^2_o) I_5 + 2 x_o (-1 +
    x_o^2) I_6}{(1+x_o^2)^2} \\
\bar{\pi}^{(3,1)} &=&  \frac{1}{2} 
\frac{2x_o(-1+x_o^2) I_5 + (1+5x_o^2) I_6}{(1+x_o^2)^2} .
\end{eqnarray} 
\end{subequations}
Note that $\bar{\pi}^{(0,2)} = - \bar{\pi}^{(2,0)}$.

The coefficients of the NLO terms appearing in the quadratic Lagrangian
(\ref{qua}), which are somewhat laborious to compute, can again be
expressed as linear combinations of $I_5$ and $I_6$, with the results
\begin{subequations}
\begin{eqnarray}
 b_1 &=& 
\frac{1}{4} \frac{(3 + 2 x_o^2) I_5 - x_o
  I_6}{1+x_o^2} \\
b_2 &=& - \frac{1}{12} \frac{x_o(1 + 4 x_o^2)
  I_5 + (7 + 10 x_o^2) I_6}{1+x_o^2} \\  c_1 &=& - \frac{1}{4} 
\frac{x_o I_5 + I_6}{1+x_o^2} \\
c_2 &=& \frac{1}{4} 
\frac{(3+4x_o^2) I_5 + x_o I_6}{1+x_o^2} \\
c_3 &=&  \frac{1}{5} 
[(3+4x_o^2) I_5 + x_o I_6] \\
c_4 &=& - \frac{1}{3} (4x_o I_5 + I_6) \\
d_1 &=& \frac{1}{2}  \frac{I_5 -
  x_o I_6}{1+x_o^2} \\
d_2 &=& - \frac{1}{2} \frac{x_o I_5 +
  I_6}{1+x_o^2} \\
d_4 &=&   I_5 .
\end{eqnarray} 
\end{subequations}

\bibliographystyle{apsrev}
\bibliography{adb}

\end{document}